\documentstyle[epsfig,prl,aps]{revtex}
\begin{document}
\draft \twocolumn[\hsize\textwidth\columnwidth\hsize\csname
@twocolumnfalse\endcsname
\title{Cosmological term, mass, and space-time symmetry $^{*}$}
\author{Irina Dymnikova
}
\address{Department of Mathematics and Computer Science,
         University of Warmia and Mazury,\\
Zolnierska 14, 10-561 Olsztyn, Poland; e-mail:
irina@matman.uwm.edu.pl}

\maketitle

\begin{abstract}
In the spherically symmetric case the requirements of regularity
of density and pressures and finiteness of the ADM mass $m$,
together with the weak energy condition, define the family of
asymptotically flat globally regular solutions to the Einstein
minimally coupled equations which includes the class of metrics
asymptotically de Sitter as $r\rightarrow 0$. A source term
connects smoothly de Sitter vacuum in the origin with the
Minkowski vacuum at infinity and corresponds to anisotropic vacuum
defined macroscopically by the algebraic structure of its
stress-energy tensor which is invariant under boosts in the radial
direction. Dependently on parameters, geometry describes vacuum
nonsingular black holes, and self-gravitating particle-like
structures whose ADM mass is related to both de Sitter vacuum
trapped in the origin and smooth breaking of space-time symmetry.
The geometry with the regular de Sitter center has been applied to
estimate geometrical limits on sizes of fundamental particles, and
to evaluate the gravito-electroweak unification scale from the
measured mass-squared differences for neutrino.

\vskip0.1in

$^{*}$ {\bf Talk at the Fourth International Conference on Physics
Beyond the Standard Model "Beyond the Desert'03", Castle Ringberg,
Germany, June 2003; to appear in "Physics Beyond the Standard
Model, BEYOND'03", Ed. H.V. Klapdor-Kleingrothaus}

\end{abstract}


 \vskip0.2in
]
\section{The Einstein Cosmological Term}
\centerline{\it  ~~~ ~ ~~  ~~ ~~   ~ ~~~~  ~~~~~~Which arrow does
fly forever?} \centerline{\it ~~ ~~  ~~ ~~   ~~~~  ~~~ ~~~   ~~An
arrow which hits the goal.} \centerline{\it ~~ ~~  ~~~~~~~~~~~~ ~~
~~ ~~ ~~~~ ~~~ ~Vladimir Nabokov} \vskip0.1in In 1917 Einstein
introduced a cosmological term into his equations describing
gravity as space-time geometry (G-field) generated by matter
$$ G_{\mu\nu}=-8\pi G T_{\mu\nu}                                          \eqno(1)
$$
to make them consistent with Mach's principle, one of his basic
motivations \cite{eins}, which reads: some matter has the property
of inertia only because there exists also some other matter in the
Universe \cite{sciama}. When Einstein found that Minkowski
geometry is the regular solution to (1) without source term
$$ G_{\mu\nu}=0                                                \eqno(2)
$$
perfectly describing inertial motion in the absence of a matter,
he modified his equations by adding the cosmological term $\Lambda
g_{\mu\nu}$ in the hope that modified equations
$$ G_{\mu\nu}+\Lambda g_{\mu\nu}=-8\pi G T_{\mu\nu}
                                                                 \eqno(3)
$$
will have reasonable regular solutions only when mattaer is
present (if matter is the source of inertia, then in case of its
absence there should not be any inertia \cite{bondi}). The
by-product of this hypothesis was the static Einstein cosmology in
which the task of $\Lambda$ was to make a universe

The story of abandoning $\Lambda$ by Einstein is typically told as
dominated by successes of FRW cosmology confirmed by Hubble's
discovery of the Universe expansion. In reality the first reason
was de Sitter solution: soon after introducing $\Lambda
g_{\mu\nu}$, in the same year 1917, de Sitter found quite
reasonable solution to the equation (3) with $T_{\mu\nu}=0$,
$$
G_{\mu\nu}+\Lambda g_{\mu\nu}=0
                                                    \eqno(4)
$$
 which made evident that a matter is not necessary to
produce the property of inertia \cite{weinberg}.

In de Sitter geometry \cite{desitter}
$$
ds^2=\biggl(1-\frac{\Lambda}{3} r^2 \biggr) dt^2 -
\frac{dr^2}{1-\frac{\Lambda}{3}r^2}- r^ d\Omega^2
                                                      \eqno(5)
$$
$\Lambda$ must be constant by virtue of the contracted Bianchi
identities
$$
G_{;\nu}^{\mu\nu}=0 ~~ ~~~\rightarrow ~~ ~~~~ \Lambda=const
                                                          \eqno(6)
$$
 It plays the role of a universal repulsion whose physical sense
remained obscure during several decades when de Sitter geometry
has been mainly used as a simple testing ground for developing the
quantum field technics in a curved space-time.

In 1965 two papers  shed some light on the physical nature of the
de Sitter geometry. In the first Sakharov suggested that
gravitational effects can dominate an equation of state at
superhigh densities and that one of possible equations of state
for superdense matter is \cite{sakharov}
$$
p=-\rho
                                                         \eqno(7)
$$
which formally corresponds to equation of state for $\Lambda
g_{\mu\nu}$ shifted to the right-hand side of the Einstein
equation (4) as some stress-energy tensor
$$
G_{\mu\nu}=- \Lambda g_{\mu\nu}=- 8\pi G T_{\mu\nu}
                                                     \eqno(8)
$$
The physical sense of this operation has been clarified in the
second paper by Gliner who identified $\Lambda g_{\mu\nu}$ on the
basis of the Petrov classification \cite{petrov}, as corresponding
to a stress-energy tensor for a vacuum defined by the algebraic
structure of its stress-energy tensor \cite{gliner}
$$
 T_{\mu\nu}=\rho_{vac} g_{\mu\nu};~~~
\rho_{vac}=(8\pi G)^{-1} \Lambda =const
                                                     \eqno(9)
$$
In quantum field theory a vacuum state which behaves like an
effective cosmological term
$$
<T_{\mu\nu}>=<\rho_{vac}> g_{\mu\nu}
                                                   \eqno(10)
$$
was first found by De Witt \cite{dewitt} (for review
\cite{adler,weinberg}).

Properties of de Sitter geometry ultimately advanced $\Lambda$ to
provide the generic reason for the Universe expansion producing a
huge growth of the scale factor \cite{us75} sufficient to explain
various puzzles of the standard big bang cosmology (for review
\cite{olive}).

The triumphs of the inflationary paradigm somehow left in shadow
the primary sense of Einstein's message of introducing $\Lambda$
as a quantity which can have something in common with inertia.

The question of possible connection between $\Lambda$ and inertia,
touches the other Einstein's profound proposal - to describe an
elementary particle by regular solution of nonlinear field
equations as "bunched field" located in the confined region where
field tension and energy are particularly high \cite{bunch}.

The possible way to such a structure of gravitational origin whose
mass is related to $\Lambda$ and whose regularity is related to
this fact, can be found in the Einstein field equations (1)  and
in the Petrov classification for $T_{\mu\nu}$. On this way one
possible answer comes from model-independent analysis of the
minimally coupled spherically symmetric Einstein equations if
certain general requirements are satisfied
\cite{me2002,me2003,stab}:

(a) regularity of density $\rho(r)$,

(b) finiteness of the ADM mass $m$ and

(c1) dominant energy condition for $T_{\mu\nu}$ either

(c2) weak energy condition for $T_{\mu\nu}$ and regularity of
pressures.

Cases (c1)-(c2) differ by behavior of the curvature scalar $R$,
which in the first case  is non-negative \cite{stab}.

Conditions (a)-(c) lead to the existence of globally regular
geometry asymptotically de Sitter as $r\rightarrow 0$, called de
Sitter-Schwarzschild geometry in case of Minkowski asymptotic at
infinity \cite{me92,me96,me2000} (for review \cite{merev}). We
applied this geometry to estimate geometrical limits on sizes of
fundamental particles in testing to which extent they can be
treated as point-like \cite{ETHZ}, and to study space-time origin
of a mass-squared difference and evaluate the gravity-electroweak
unification scale from the mass squared differences for neutrino
\cite{neutrino,precise}.

This talk is organized as follows. In Section II we present
conditions leading to the existence of geometry with the regular
de Sitter center. In Section III we outline the structure of a
source term for this geometry and in Section IV geometry itself.
Sections V and VI are devoted to tests: estimates of limits on
geometrical sizes of leptons and extraction of the
gravito-electroweak scale from the data on mass squared
differences for neutrino.

\section{Regular De Sitter Center}

We start from the Einstein field equation (1) without cosmological
term.

The standard form for a static spherically symmetric line element
reads \cite{tolman}
$$      ds^2 = e^{\mu(r)}dt^2 - e^{\nu(r)} dr^2 - r^2 d\Omega^2
                                                            \eqno(11)
 $$
where $d\Omega^2$ is the metric on a unit 2-sphere. The metric
functions satisfy the Einstein equations
$$     8\pi G T_t^t = 8\pi G\rho(r)= e^{-\nu}\biggl(\frac{{\nu}^{\prime}}{r}
-\frac{1}{r^2}\biggr)     +\frac{1}{r^2}
                                                                    \eqno(12.1)
$$
$$
8\pi G T_r^r =-8\pi G p_r(r)= -e^{-\nu}
\biggl(\frac{{\mu}^{\prime}}{r} +\frac{1}{r^2}\biggr)
+\frac{1}{r^2}
                                                                      \eqno(12.2)
$$
$$
8\pi G T_{\theta}^{\theta}=8\pi G T_{\phi}^{\phi}=-8\pi G
p_{\perp}(r)=$$
$$-e^{-\nu}\biggl(\frac{{{\mu}^{\prime\prime}}}{2}
 +\frac{{{\mu}^{\prime}}^2}{4}
 +\frac{({{\mu}^{\prime}-{\nu}^{\prime}})}{2r}-\frac{{\mu}^{\prime}
     {\nu}^{\prime}}{4}\biggr)
                                                                         \eqno(12.3)
$$
Here $\rho(r)=T^t_t$ is the energy density (we adopted $c=1$ for
simplicity), $p_r(r)=-T^r_r$ is the radial pressure, and
$p_{\perp}(r)=-T_{\theta}^{\theta}=-T_{\phi}^{\phi}$ is the
tangential pressure for anisotropic perfect fluid \cite{tolman}.
The prime denotes differentiation  with respect to $r$.
Integration of Eq.(12.1) gives
$$
e^{-\nu(r)}=1-\frac{2GM(r)}{r};~~M(r) =4\pi\int_0^r{\rho(x)x^2dx}
                                                                    \eqno(13)
$$
whose asymptotic for large $r$ is $e^{-\nu}=1-{2Gm}/{r}$, with the
mass parameter $m$ given by
 $$
 m=4\pi\int_0^{\infty}{\rho(r) r^2 dr}
                                                                  \eqno(14)
$$
The dominant energy condition $T^{00}\geq|T^{ab}|$ for each
$a,b=1,2,3$, which holds if and only if \cite{HE}
      $$\rho\geq0;~~~~-\rho\leq p_k\leq \rho;~~~~k=1,2,3
                                                                    \eqno(15)
$$
implies that the local energy density is non-negative for any
observer in his local frame, and each principal pressure never
exceeds the energy density. In the limit $r\rightarrow\infty$ the
condition of finiteness of the mass (14) requires density profile
$\rho(r)$ to vanish at infinity quicker than $r^{-3}$, and then,
in the case (c1) the dominant energy condition (15) requires both
radial and tangential pressures to vanish as $r\rightarrow\infty$.
Then $\mu^{\prime}=0$ and $\mu=$const at infinity. Rescaling the
time coordinate allows one to put the standard boundary condition
$\mu\rightarrow 0$ as $r\rightarrow \infty$ \cite{wald} which
leads to asymptotic flatness needed to identify (14) as the ADM
mass \cite{wald}. In the case (c2) we postulate regularity of
pressures including vanishing of $p_r$ at infinity sufficient to
get $\mu^{\prime}=0$ needed for asymptotic flatness.

From Eqs.(12) hydrodynamic equations follow \cite{oppi,apj,me2002}
$$
p_{\perp}=p_r+\frac{r}{2}p_r^{\prime}+(\rho+p_r)\frac{G M(r)
     +4\pi G r^3 p_r}{2(r-2G M(r))}
                                                                    \eqno(16)
$$
$$
p_r+\rho=\frac{1}{8\pi G}\frac{e^{-\nu}}{r}
 (\nu^{\prime}+\mu^{\prime})
                                                                 \eqno(17)
$$
which define, by imposing requirements (a)-(c), asymptotic
behavior of a mass function $M(r)$ at approaching the regular
center:
$$
M(r) ~~ \rightarrow \frac{4\pi}{3}\rho(0) r^3 ~~\quad {as} ~~~
r\rightarrow 0
                                                                           \eqno(18)
$$
Eq.(17) leads to $\nu^{\prime}+\mu^{\prime}=0$ and
$\nu+\mu=\mu(0)$ at $r=0$ with $\mu(0)$ playing the role of the
family parameter \cite{me2002}.

The weak energy condition, $T_{\mu\nu}\xi^{\mu}\xi^{\nu}\geq 0$
for any time-like vector $\xi^{\mu}$, which means non-negative
density for each observer on a time-like curve, is contained in
the dominant energy condition and satisfied if and only if
$$
\rho\geq 0; \rho + p_k \geq 0, k=1,2,3
                                                                             \eqno(19)
$$
By Eq.(17) it demands $\mu^{\prime}+\nu^{\prime}\geq 0$ everywhere
\cite{me2002}. A function $\mu(r)+\nu(r)$ is  growing from
$\mu+\nu=\mu(0)$ at $r=0$ to $\mu+\nu=0$ at $r\rightarrow\infty$,
which leads to $\mu(0)\leq 0$ \cite{me2002}. \footnote{The well
known example of solution from this family is boson stars
\cite{boson} (for review \cite{Mielke}).}

The range of family parameter $\mu(0)$ includes the value
$\mu(0)=0$. In this case the function $\nu(r)+\mu(r)$ is zero at
$r=0$ and at $r\rightarrow\infty$, its derivative is non-negative,
hence  $\nu(r)=-\mu(r)$ everywhere \cite{me2002}, and the metric
is
$$
ds^2=g(r)dt^2-\frac{dr^2}{g(r)}-r^2 d\Omega^2
                                                                 \eqno(20)
$$
with $$ g(r) = 1 -\frac{2GM(r)}{r}; ~~~~~M(r)=4\pi
\int_0^r{\rho(x)x^2 dx}
                                                                    \eqno(21)
$$
The metric (20)  has Schwarzschild asymptotic as $r\rightarrow
\infty$
$$
ds^2=\biggl(1-\frac{r_g}{r}\biggr)-
 \frac{dr^2}{\biggl(1-\frac{r_g}{r}\biggr)}-r^2d\Omega^2; ~~r_g=2Gm
                                                                  \eqno(22)
$$
The weak energy condition defines also equation of state and thus
asymptotic behavior as $r\rightarrow 0$ \cite{me2002}: In the
limit $r\rightarrow 0$ the equation of state becomes $p=-\rho$,
which gives de Sitter asymptotic as $r\rightarrow 0$
$$
T_{\mu\nu}=\rho_0 g_{\mu\nu}; ~~
ds^2=\biggl(1-\frac{\Lambda}{3}{r_2}\biggr)dt^2
    -\frac{dr^2}{\biggl(1-\frac{\Lambda}{3}{r_2}\biggr)}-r^2d\Omega^2
                                                                        \eqno(23)
$$
with a cosmological constant
$$
\Lambda =8\pi G \rho_0
                                                                        \eqno(24)
$$
where $\rho_0=\rho(r\rightarrow 0)$.

We see that $\Lambda$ has appeared at the origin although it was
not present in the basic equations.

The weak energy condition $p_{\perp}+\rho\geq 0$ demands monotonic
decreasing of a density profile. The simple analysis similar to
previous shows that the metric (20)-(21) has not more than two
horizons \cite{me2002}. \vskip0.1in

{\bf Summary}

 Requirements (a)-(c) imposed on the Einstein equations
{\it without cosmological term} lead to the existence of the class
of metrics asymptotically de Sitter as $r\rightarrow 0$ and
asymptotically Schwarzschild as $r\rightarrow\infty$, with
monotonically decreasing density profile, $\rho^{\prime} \leq 0$,
 with the metric function which smoothly evolves between
$$
1-\frac{\Lambda}{3} r^2 ~~~\leftarrow ~~~~g(r)~~~\rightarrow  ~~
~1-\frac{2Gm}{r}
$$
and which has  not more than two horizons.

\section{Structure of a Source Term}

For the class of metrics (20) a source term has the algebraic
structure \cite{me92}
$$
T_t^t=T_r^r;~~~T_{\theta}^{\theta}=T_{\phi}^{\phi}
                                                       \eqno(25)
$$
and the equation of state
$$
p_r=-\rho;~~~p_{\perp}=-\rho-\frac{r}{2}\rho^{\prime}
                                                                \eqno(26)
$$
In the Petrov classification scheme \cite{petrov} stress-energy
tensors are classified on the basis of their algebraic structure.
When the elementary divisors of the matrix $T_{\mu\nu}-\lambda
g_{\mu\nu}$ are real,  the eigenvectors of $T_{\mu\nu}$ are
non-isotropic and form a co-moving reference frame with a
time-like vector representing a velocity. A co-moving reference
frame is thus defined uniquely if and only if none of the
space-like eigenvalues $\lambda_{a}$ ($a=1,2,3$) coincides with
the time-like eigenvalue $\lambda_0$. Otherwise there exists an
infinite set of co-moving reference frames, which makes impossible
to define a velocity in principle. This classification provides
general macroscopic definition of a vacuum by symmetry of its
stress-energy tensor.

The stress-energy tensor (9) corresponding to the Einstein
cosmological term has the structure [(IIII)] in the Petrov
classification (all eigenvalues equal), and is identified as a
vacuum tensor due to the absence of a preferred co-moving
reference frame \cite{gliner}.

 A stress-energy tensor (25) with the structure [(II)(II)], has an infinite set
of co-moving reference frames, since it is invariant under
rotations in the $(r,t)$ plane. Therefore an observer moving
through a medium (25) cannot in principle measure the radial
component of his velocity. The stress-energy tensor with the
algebraic structure (25) is identified thus as describing a
spherically symmetric anisotropic vacuum with variable density and
pressures, $T_{\mu\nu}^{vac}$, invariant under boosts in the
radial direction \cite{me92}.

 Any source terms for the class of metrics (20) evolves smoothly and monotonically
from de Sitter vacuum in the center to Minkowski vacuum at
infinity
$$
\quad {\it de ~Sitter} ~~~ \Lambda \delta^{\mu}_{\nu}~ ~~
 ~ \rightarrow ~~~~8\pi G T^{\mu}_{\nu}  ~~ ~\rightarrow ~~0 ~ \quad {\it Minkowski}
$$
ADM mass $m$ responsible for geometry and identified as a
gravitational mass by asymptotic behavior of the metric at
infinity, is equal, by the equivalence principle, to inertial mass
which is thus related to both de Sitter vacuum trapped inside an
object, $\Lambda = 8\pi G \rho_0$, and to reducing of symmetry of
a source term (25) from the full Lorentz group in the origin to
the radial boosts only. Symmetry of space-time generated by such a
source, is
$$
r\rightarrow 0 \quad {\it de~ Sitter~ group}~ ~~
 ~~~~~~~~~r\rightarrow \infty \quad{\it Poincare~ group}
$$
\vskip0.1in

This approach can be extended to the case of de Sitter asymptotic
at infinity, with $\lambda < \Lambda$. A density component of a
source term is taken as $T^0_0=\rho(r) + (8\pi G)^{-1}\lambda$,
and the metric function $g(r)$ in (20) is then given by
\cite{us97}
$$
g(r)=1-\frac{2GM(r)}{r} -\frac{\lambda r^2}{3}
                                                    \eqno(27)
$$
This metric is asymptotically de Sitter at both origin and
infinity and has not more than three horizons for the case of two
Lambda scales \cite{us03}. A source term evolves smoothly and
monotonically between two de Sitter vacua with different values of
cosmological constant. \vskip0.1in $ ~~~ ~~ ~~ ~~(\Lambda +
\lambda) \delta^{\mu}_{\nu} ~~~ ~\rightarrow~~~8\pi G
T^{\mu}_{\nu} ~~~\rightarrow ~~~
 ~~\lambda \delta ^{\mu}_{\nu} $  $ \quad {\it De ~Sitter ~vacuum}~
 ~~~~~~~~~~~~~ ~~ ~~~ ~~~ ~~ ~~~~\quad {\it de ~Sitter~ vacuum} $
\vskip0.1in

This makes it possible to interpret $T_{\mu\nu}^{vac}$ with the
algebraic structure (25) and  such asymptotic behavior  as
corresponding to extension of the algebraic structure of the
cosmological term from $\Lambda g_{\mu\nu}$, with $\Lambda$=const,
to an $r-$dependent cosmological term
$$
\Lambda_{\mu\nu}=8\pi G T_{\mu\nu}^{vac} \eqno(28)
$$
evolving  monotonically from $ \Lambda_{\mu\nu}=\Lambda
g_{\mu\nu}$ at $r=0$ to $\Lambda_{\mu\nu}=\lambda g_{\mu\nu}$ as
$r\rightarrow\infty$ \cite{me2000}.

The advantage of such an extension is that a scalar $\Lambda$
describing vacuum energy density as $\rho_{vac}=8\pi G \Lambda$
with $\rho_{vac}=$const by virtue of the contracted Bianchi
identities, becomes explicite related to the appropriate
component, $\Lambda^0_0$, of an appropriate stress-energy tensor,
whose vacuum properties follow from its symmetry, and whose
variability follows just from the contracted Bianchi identities
which give $\Lambda^{\mu}_{\nu;\mu}=0$.

Shifting vacuum tensor $8\pi G T^{\mu}_{\nu}$ to the left hand
side of Einstein equation (1) we obtain
$$
G^{\mu}_{\nu}+\Lambda^{\mu}_{\nu}=0 \eqno(29)
$$
In $\Lambda^{\mu}_{\nu}$ geometry a mass $m$ is directly connected
to cosmological term $\Lambda_{\mu\nu}$ by the ADM formula (14)
which in this case reads
$$
m=(2 G)^{-1} \int_0^{\infty}{(\Lambda_t^t(r)-\lambda) r^2 dr}
                                                                           \eqno(30)
$$
where $\lambda$ is the asymptotic value of $\Lambda^t_t$ at
infinity.

Let us note that this observation does not depend on
identification of a vacuum tensor of the algebraic structure (25)
as associated with a variable cosmological term (28). Any
stress-energy tensor from the considered class generates
space-time invariant under de Sitter group in the limit
$r\rightarrow 0$. And for any metric from this class the standard
formula (14) relates the ADM mass $m$ to both de Sitter vacuum
trapped in the origin and breaking of space-time symmetry
\cite{me2002}.

\section{ De Sitter-Schwarzschild Geometry}

Here we discuss de Sitter-Schwarzschild geometry (21) with
Minkowsi asymptotic at infinity.

The key point of de Sitter-Schwarzschild geometry is the existence
of two horizons, a black hole event horizon $r_{+}$ and an
internal horizon $r_{-}$. A critical value of a mass exists,
$m_{crit}$, at which the horizons come together and which puts a
lower limit on a black hole mass \cite{me96}. De
Sitter-Schwarzschild configurations are shown in Fig.1.

\begin{figure}[h]
\vspace{-8.0mm}
\begin{center}
\epsfig{file=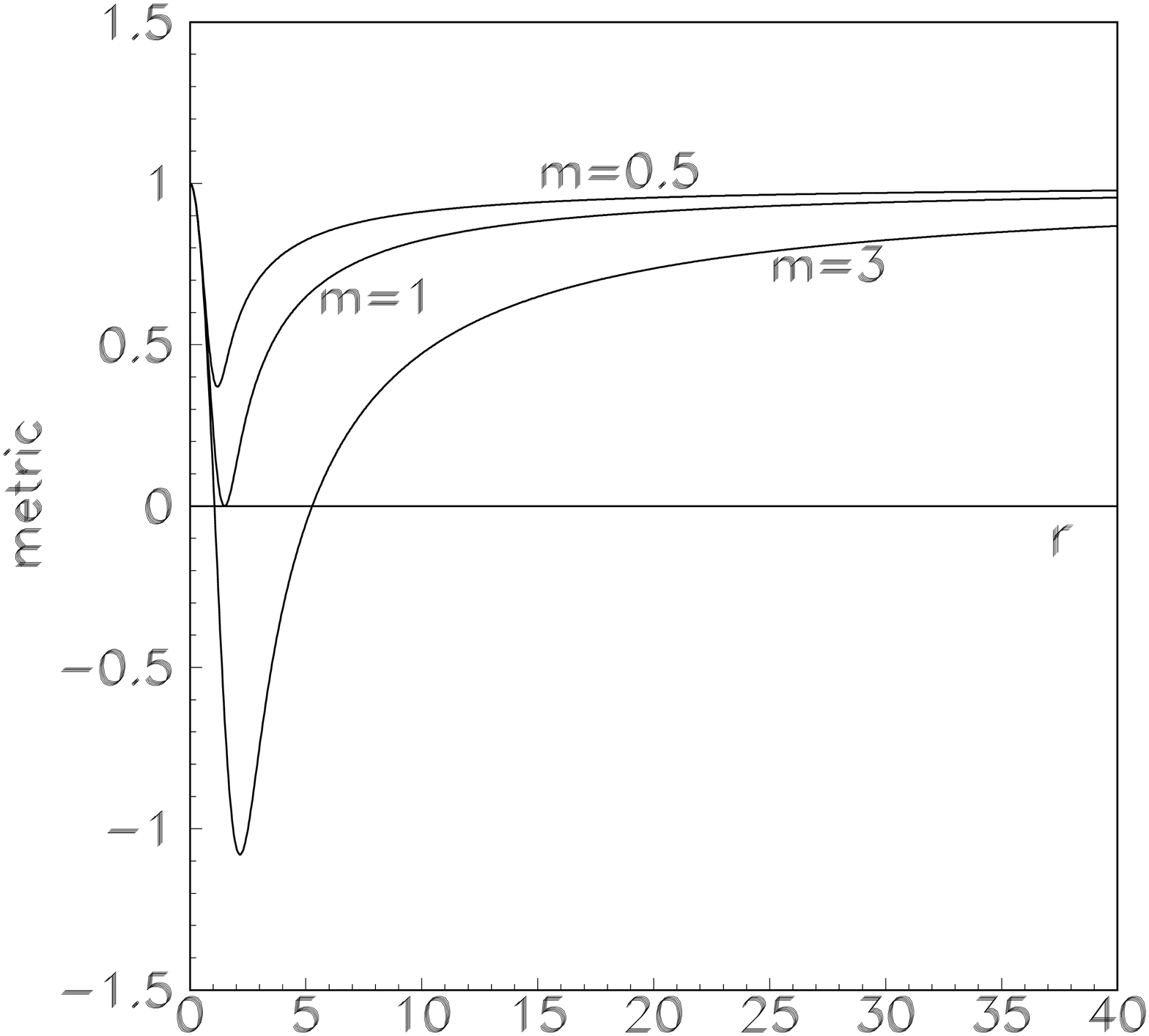,width=8.0cm,height=4.5cm}
\end{center}
\caption{De Sitter-Schwarzschild configurations. Mass parameter is
normalized to $m_{crit}$.} \label{fig.1}
\end{figure}
For the case of the density profile \cite{me92}
$$
\rho(r)=\rho_0 e^{-r^3/r_0^2 r_g}; ~~r_0^2=3/\Lambda;~~r_g=2Gm
                                                              \eqno(31)
$$
modelling semiclassically \cite{igor} vacuum polarization in the
spherically symmetric gravitational field \cite{me96}
$$
m_{crit}\simeq{0.3 m_{Pl}\sqrt{\rho_{Pl}/\rho_0}}
                                                                   \eqno(32)
$$
For $m\geq m_{crit}$ de Sitter-Schwarzschild geometry describes
the vacuum nonsingular black hole \cite{me92}, whose future and
past singularities are replaced with regular cores asymptotically
de Sitter as $r\rightarrow 0$.

It emits Hawking radiation from both black hole and cosmological
horizon with the Gibbons-Hawking temperature $T=\hbar \kappa(2\pi
kc)^{-1}$\cite{gh} where $\kappa$ is the surface gravity. The form
of the temperature-mass diagram is generic for de
Sitter-Schwarzschild geometry. The temperature on the BH horizon
drops to zero as $m\rightarrow {m_{crit}}$, while the
Schwarzschild asymptotic requires $T_{+}\rightarrow 0$ as
$m\rightarrow\infty$. As a result the temperature-mass curve has a
maximum between $m_{crit}$ and $m\rightarrow\infty$, where a
specific heat is broken and changes sign testifying to a
second-order phase transition in the course of Hawking evaporation
and suggesting restoration of space-time symmetry to the de Sitter
group in the origin \cite{me97}.

For masses $m<m_{crit}$ de Sitter-Schwarzschild geometry describes
a self-gravitating particle-like vacuum structure without
horizons, globally regular and globally neutral. It is plotted in
Fig. 2 for the density profile (31).

\begin{figure}[h]
\vspace{-8.0mm}
\begin{center}
\epsfig{file=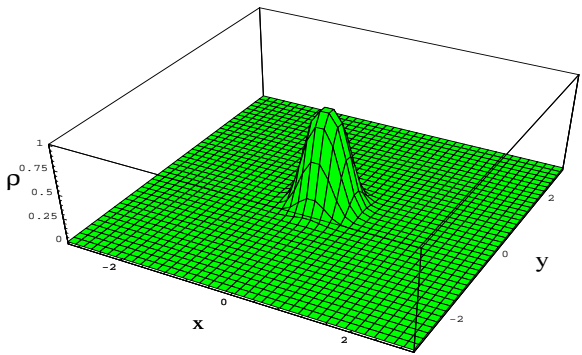,width=8.0cm,height=5.0cm}
\end{center}
\caption{ G-lump in the case $r_g=0.1r_0$ ($m\simeq{0.06
m_{crit}})$. } \label{fig.2}
\end{figure}

It resembles Coleman's lumps - non-singular, non-dissipative
solutions of finite energy, holding themselves together by their
own self-interaction \cite{lump}. G-lump is the regular solution
to the Einstein equations, perfectly localized (see Fig. 2) in a
region where field tension and energy are particularly high (this
is the region of the former singularity). It holds itself together
by gravity due to balance between gravitational attraction outside
and gravitational repulsion inside of zero-gravity surface $r=r_c$
beyond which the strong energy condition of singularities theorems
\cite{HE}, $(T_{\mu\nu}-T g_{\mu\nu}/2)\xi^{\mu}\xi^{\nu})\geq 0$,
is violated \cite{me96}. The surface of zero gravity is depicted
in Fig. 3 together with horizons and with the surface $r=r_s$ of
zero scalar curvature $R(r_s)=0$ which represents the
characteristic curvature size in the de Sitter-Schwarzschild
geometry in the case (c2).
\begin{figure}[h]
\vspace{-8.0mm}
 \begin{center}
\epsfig{file=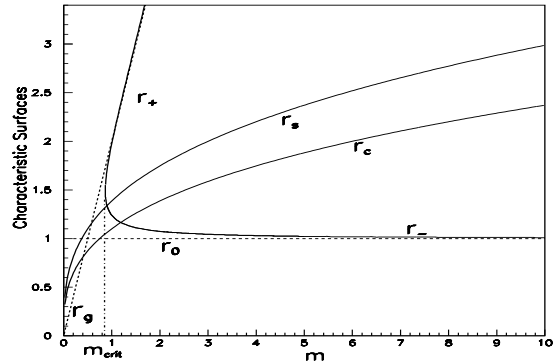,width=8.0cm,height=5.5cm}
\end{center}
\caption{ Horizons of $\Lambda$BH, surfaces $r=r_s$ and $r=r_c$. }
\label{fig.3}
\end{figure}

For the density profile (31) the characteristic size $r_s$ is
given by \cite{me96}
$$
r_s=\biggl(\frac{4}{3}r_0^2 r_g\biggr)^{1/3}
 =\biggl(\frac{m}{\pi\rho_0}\biggr)^{1/3}
                                                                       \eqno(33)
$$

The  mass of  G-lump is directly connected to a de Sitter vacuum
trapped in its center  and to breaking of space-time symmetry from
the de Sitter group in the origin to the Poincare group at
infinity.

This picture conforms with the basic idea of the Higgs mechanism
for generation of mass via spontaneous breaking of symmetry of a
scalar field vacuum from a false vacuum to a true vacuum state. In
both cases de Sitter vacuum is involved and vacuum symmetry is
broken.

The difference is that the gravitational potential $g(r)$ (shown
in Fig. 4) is generic, and the de Sitter vacuum supplies a
particle with mass via smooth breaking of space-time symmetry.
\begin{figure}
\vspace{-8.0mm}
\begin{center}
\epsfig{file=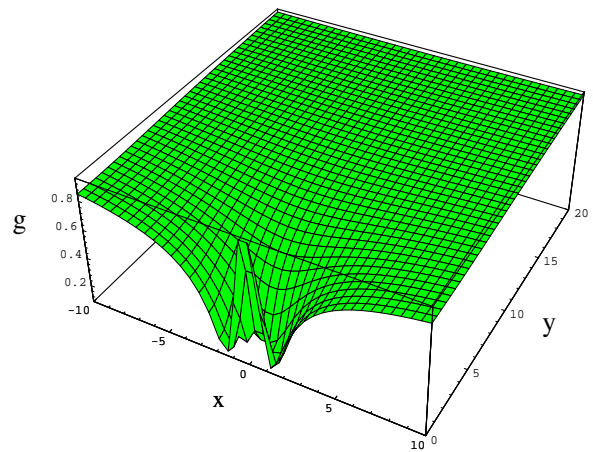,width=8.0cm,height=6.5cm}
\end{center}
\caption{The gravitational potential $g(r)$ for the case of G-lump
with the mass a little bit less than $m_{crit}$. } \label{fig.4}
\end{figure}
\vskip0.1in

{\bf Summary}

In de Sitter-Schwarzschild geometry de Sitter vacuum is involved
generically in mass generation via smooth breaking of space-time
symmetry from de Sitter group in the origin to the Poincare group
at infinity.

\section{Test 1:  Limits on Sizes of Fundamental Particles}

To test de Sitter-Schwarzschild geometry we estimate geometrical
limits on sizes of fundamental particles by characteristic
geometrical size given by curvature radius $r_s$ (see Fig. 3).
This implies rather natural assumption that whichever would be
particular mechanism involving de Sitter vacuum in mass
generation, a fundamental particle may have an internal vacuum
core related to its mass and a geometrical size defined by
gravity. Geometrical size of an object with the de Sitter vacuum
in the origin at the background of the Minkowski vacuum at
infinity, can be approximated by de Sitter-Schwarzschild geometry.
Characteristic size in this geometry depends on vacuum density at
$r=0$ and actually presents a modification of the Schwarzschild
radius $r_g$ to the case when singularity is replaced with the de
Sitter vacuum. The resulting difference in sizes is quite
impressive:  for elementary particles the Schwarzschild radius
give sizes which are many orders of magnitude less than $l_{Pl}$;
the characteristic de Sitter-Schwarzschild radius $r_s$ gives
sizes close to the experimental upper limits (e.g.,
$r_s\sim{10^{-18}}$ cm for the electron getting its mass from the
vacuum at the electroweak scale). In Fig. 5\cite{ETHZ} geometrical
sizes $r_s$ plotted by dark triangles, are compared with
electromagnetic (EM) and electroweak (EW) experimental limits.
\begin{figure}[h]
\vspace{-8.0mm}
\begin{center}
\epsfig{file=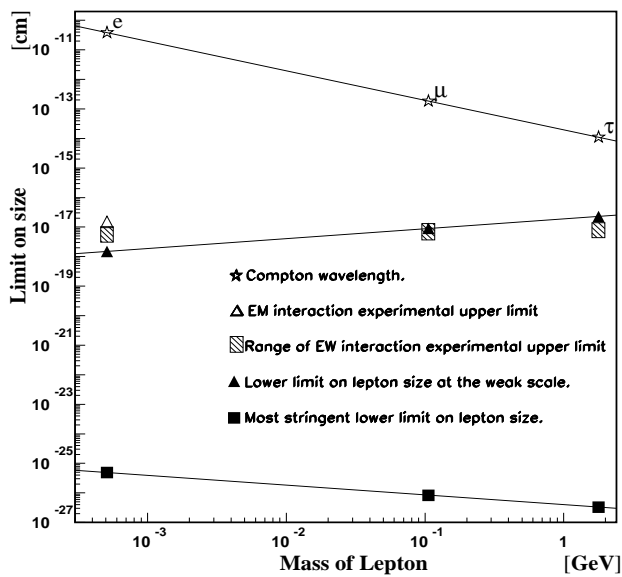,width=9.0cm,height=8.5cm}
\end{center}
\caption{Characteristic sizes for leptons \protect\cite{ETHZ}. }
\label{fig.5}
\end{figure}

In Fig. 5 we show by stars quantum limits given by Compton wave
length, by white triangles electromagnetic experimental upper
limits coming mainly from reaction $e^{+}e^{-}--\rightarrow
\gamma\gamma(\gamma)$ \cite{L3}, by shaded squares experimental
electro-weak limits \cite{ew}, by dark triangles geometrical
limits on sizes calculated from Eq.(33) with $\rho_0$ of the
electro-weak scale 246 GeV, and by dark squares the most stringent
lower limits on sizes of particles as extended objects. This last
limit is calculated by taking into account that in the case when
de Sitter vacuum is involved in mass generation, quantum region of
a particle localization $\lambda_C$ must fit within a causally
connected region confined by the de Sitter horizon $r_0$. The
scale $r_0$ is characteristic de Sitter radius related to density
$\rho_0$ by
$$
r_0^2=\frac{3c^2}{8\pi G \rho_0}
                                               \eqno(34)
$$
The requirement $\lambda_C\leq r_0$ gives the limiting scale for a
vacuum density $\rho_0$ related to a given mass $m$
$$
\rho_0 \leq \frac{3}{8\pi}\biggl(\frac{m}{m_{Pl}}\biggr)^2
\rho_{Pl}
                                                                         \eqno(35)
$$
This condition connects a mass of a quantum object $m$ with the
scale for vacuum density $\rho_0$ at which this mass could be
generated in principle
 provided that a mechanism of generation involves de Sitter vacuum.

In Fig. 3 and Fig. 5 theoretical estimates are calculated with
using the density profile (31), but the  results would not change
drastically for different profiles, since a characteristic length
scale in any spherical geometry involving de Sitter center,  is
$r_*\sim{(r_0^2 r_g)^{1/3}}$ \cite{werner}.

 Let us compare characteristic sizes for an electron, its Compton wavelength,
classical and Schwarzschild radius
$$
\lambda_C \simeq {3.9 \times 10^{-11}~ cm}; ~ r_{class}\simeq {2.8
\times 10^{-13}}~cm;
 $$
 $$
~r_g\simeq{10^{-57}~cm}
$$
with lower limits on geometrical sizes for the case when de Sitter
vacuum is involved on the electro-weak scale, on
gravito-electroweak scale of order of several TeV (see next
Section) and at the most stringent scale (35)
$$
 r_{EW}\simeq{1.5 \times 10^{-18}~cm};
~r_{GEW}\simeq {2\times 10^{-23} ~cm};
$$
 $$
~ r_{lowest}\simeq{5 \times 10^{-26}~cm}
$$
We see that numbers given by de Sitter-Schwarzschild geometry are
much bigger that the Planck scale $l_{Pl}\sim{10^{-33}}$ cm, which
justifies application of classical General Relativity for
estimation of geometrical sizes of quantum particles.

\section{Test 2:  Space-time symmetry as origin of mass-square difference}

In this Section we outline the papers \cite{neutrino,precise}.

If in the interaction region where particles are created, the
interaction vertex is gravito-electroweak, gravity is involved
essentially, so geometry around the vertex is not Minkowski any
more, and mass is not Casimir invariant of the Poincare group.  If
density in the vertex is limited, mass of a particle is finite and
some of conditions (c) holds, i.e. energy density is non-negative,
then geometry around the vertex can be de Sitter. If a false
vacuum is  somehow involved (for example via Higgs mechanism),
then geometry  in the interaction region is de Sitter.

If de Sitter group is the space-time symmetry group induced around
the gravito-electroweak vertex, then particles participating in
the vertex are described by the eigenstates of Casimir operators
of the de Sitter group. Their further evolution in Minkowski
background requires further symmetry change. We suggest that the
"flavor" can emerge due to change in symmetry of space-time
between interaction region and propagation region, in particular
from de Sitter group around the vertex to the Poincare group
outside \cite{neutrino}. This point needs further detailed
analysis which is in progress now.

What we can do immediately is to calculate an eigenstate of the
first Casimir invariant of the de Sitter group which relates a
mass with the vacuum density $\rho_0$ at the scale of unification.

 It reads \cite{fg}
$$
I_1 = \Pi_\mu \Pi^\mu - \frac{1}{2 r_0^2} J_{\mu\nu} J^{\mu\nu}
                                                                   \eqno(36)
$$
with
$$
\Pi_\mu = \left(  1+ \frac{r^2-c^2 t^2}{4 r_0^2}\right)P_\mu
 + \frac{1}{2 r_0^2} {x^\nu} J_{\mu\nu}
                                                                 \eqno(37)
$$
In the interaction region $r^2-c^2 t^2 \ll r_0^2$ (to be confirmed
below), and the operator $I_1$ is approximated by
$$
I_1 \approx  P_\mu P^\mu - \frac{1}{r_0^2}  \left({\bf J}^2-{\ bf
K}^2\right) \eqno(38)
$$
where (details in \cite{neutrino}) $ J_{ij}=-J_{ji}=
\epsilon_{ijk} J_k$ and $J_{i 0}=-J_{0 i} = - K_i$, with each of
the $i,j,k$ taking the values $1,2,3$. The ${\bf J}$ are then
generators of Lorentz rotation and $\bf K$ are generators of
Lorentz boosts.

This gives
$$
I_1 \approx  P_\mu P^\mu  - \frac{\hbar^2}{2 r_0^2}{\bf \sigma}^2
                                                                    \eqno(38)
$$
Its eigenvalues (degenerated) are:
$$
I_1^\prime = \mu^2 c^2 - \frac{3\hbar^2}{2 r_0^2}
                                                                   \eqno(39)
$$
where $\mu^2 c^2$ is the eigenvalue of the first Casimir invariant
for the Poincare group.

We see that in the de Sitter geometry the mass eigenvalues depend
on the vacuum density $\rho_0$ responsible for geometry. This
allows for negative mass-square for sub-eV particles if
gravito-electroweak unification occurs at TeV scales
\cite{precise}. This also might offer a natural explanation for
anomalous results known as "negative mass squared problem" for
${\nu}_e$ \cite{nms}.

De Sitter symmetry in the gravito-electroweak vertex gives the
characteristic mass-square scale
$$
\Delta m^2 = \frac{3\hbar^2}{2 c^2 r_0^2}
                                                            \eqno(40)
$$
Connecting $r_0$ with the unification scale $M_{unif}$ through
$\rho_0/\rho_{Pl}\sim{(M_{unif}/M_{Pl})^4}$, we get
 $$
M_{unif} \sim{ \left[ \frac{1}{4 \pi} \left( \frac{\Delta
m^2}{M_{Pl}^2} \right) \right]^{1/4} M_{Pl}}
                                                              \eqno(41)
$$
which allows us to read off a unification scale from the neutrino
mass-square data \cite{neutrino}.

The atmospheric neutrino data \cite{pakvasa}
$$
\delta m^2_{ATM}=2.5 \times 10^{-3} eV^2
$$
 give for the unification scale
$$
M_{unif} \sim{ 16~~ TeV}
                                                       \eqno(42)
$$
and mass-squared difference from solar neutrino data
\cite{pakvasa}
$$
\delta m^2_{ATM}=2.5 \times 10^{-3} eV^2
$$
gives
$$
M_{unif} \sim {6.5~~ TeV}
                                                       \eqno(43)
$$
 These correspond, respectively, to $r_0\sim {4  \times 10^{-4}}$ cm,
and $r_0 \sim{ 2\times 10^{-3}}$ cm. This justifies accuracy of
calculations: for a particle with mass $<m>_{\nu_e}=0.39$ eV,
characteristic curvature size is $r_s\sim{10^{-23}}$ cm and the
Compton wavelength $\lambda_C\sim{10^{-5}}$ cm (for curiosity, the
Schwarzschild radius is $r_g\sim{10^{-63}}$ cm).

 \section{Discussion}

The main point outlined here is the existence of the class of
globally regular solutions to the minimally coupled GR equations
with a source term of the algebraic structure (25) interpreted as
spherically symmetric anisotropic vacuum with variable density and
pressures $T_{\mu\nu}^{vac}$ associated with a time-dependent and
spatially inhomogeneous cosmological term $\Lambda_{\mu\nu} =8\pi
GT_{\mu\nu}^{vac}$, whose asymptotic behavior in the origin,
dictated by the weak energy condition, is the Einstein
cosmological term $\Lambda g_{\mu\nu}$.

De Sitter-Schwarzschild geometry  describes generic properties of
any configuration satisfying (25) and requirements (a)-(c),
 obligatory for any particular model in the same sense
as de Sitter geometry is  obligatory for any matter source
satisfying (9).

In de Sitter-Schwarzschild geometry space-time symmetry changes
smoothly from de Sitter group at the center to the Poincare group
at infinity, and the standard formula for the ADM mass $m$ relates
it (generically, since a matter source can be any from considered
class) to both de Sitter vacuum trapped inside an object and
breaking of space-time symmetry.

Applying this geometry to estimating geometrical limits on lepton
sizes, we {\it do not model} a lepton by G-lump, but only use de
Sitter-Schwarzschild geometry as a proper instrument to evaluate
the geometrical size of an object whose mass has something in
common with the de Sitter vacuum in the origin and Minkowski
vacuum at infinity.

Application of classical geometry to estimation of sizes of
quantum particles is justified by that estimates give numbers
which are very orders of magnitude above the Planck scale so that
the manifold is perfectly smooth and quantization of gravity is
not relevant.

For leptons this is rather rough estimate since more precise
geometry is needed which takes into account charge and rotation.
We are working on such a geometry.

Applying geometry with the regular de Sitter center to the problem
of neutrino mass-square difference we do not apply it literally.
The question we address is what is the  space-time symmetry group
around a gravito-electroweak vertex. In answering this question we
base on a view of a vertex as gravito-electroweak, so that gravity
is involved essentially and geometry cannot be Minkowski any more.
With de Sitter symmetry group in the interaction region, neutrinos
are described by eigenstates of the Casimir operators in de Sitter
geometry.
 We find a relation between mass-squared differences for neutrinos
and the unification scale for gravity-electroweak vertex, which
predicts a  TeV scale for unification.

  It is clear that the idea of de Sitter group in the interaction region
 is applicable  to particles besides neutrinos \cite{neutrino}.
Actually it is simple and general enough to be applicable to {\em
all\/} particles that participate in interactions where de Sitter
vacuum is involved.

\section{Acknowledgment}
This work was supported by the Polish Committee for Scientific
Research through grant No. 5P03D.007.20.

\end{document}